# Observation of Iso-Symmetric Structural and Lifshitz Transitions in Quasi-one-dimensional CrNbSe$_5$

Mingyu Xu[1], Peng Cheng[1], Shuyuan Huyan[2,3], Wenli Bi[4], Su-Yang Xu[5], Sergey L. Bud'ko[2,3], Paul C. Canfield[2,3], Weiwei Xie[1*]

1. Department of Chemistry, Michigan State University, East Lansing, MI 48824, USA
2. Ames National Laboratory, Ames, IA 50011, USA
3. Department of Physics and Astronomy, Iowa State University, Ames, IA 50011, USA
4. SmartState Center for Experimental Nanoscale Physics, Department of Physics and Astronomy, University of South Carolina, Columbia, SC 29208, USA
5. Department of Chemistry and Chemical Biology, Harvard University, Cambridge, MA 02138, USA

*Corresponding author: Dr. Weiwei Xie (xieweiwe@msu.edu)

*Abstract*

Chalcogenides-rich transition metal compounds host a rich landscape of emergent quantum phenomena that are intimately governed by their quasi-one-dimensional chemical-bonding frameworks and their response to external perturbations such as pressure. Here, we report a pressure-induced iso-symmetric structural transition in the quasi-one-dimensional compound CrNbSe$_5$, in which the electronic ground state is controlled not by symmetry breaking but by a continuous reorganization of local bonding interactions. Applied pressure reversibly tunes CrNbSe$_5$ between semiconducting and semimetallic states, enabling access to low- and high-carrier electronic regimes through direct modulation of metal–chalcogen bonding. High-pressure single-crystal X-ray diffraction directly resolves the evolution of Cr-Se and Nb-Se bond distances, coordination polyhedra, and connectivity, revealing a fully reversible semimetal-semiconductor-semimetal transition driven by gradual yet cooperative bond rearrangements within a preserved crystallographic symmetry. In contrast to chemical substitution, which irreversibly alters composition and introduces disorder, pressure acts as a clean, continuous control parameter that reshapes the bonding landscape without disrupting structural symmetry. These results establish CrNbSe$_5$ as a model system for electronically driven phase switching via tunable chemical bonding, highlighting iso-symmetric bond reorganization as a powerful design principle for pressure-controlled electronic and spintronic functionalities.

# Introduction

Low-dimensional materials, particularly quasi-one-dimensional (1D) and two-dimensional (2D) van der Waals (vdW) systems[1–3], display a rich landscape of emergent electronic, optical, and magnetic properties owing to the reduced spatial dimensionality and enhanced electronic correlations[4–9]. Among these, the transition metal trichalcogenides (TMTCs, $MX_3$; M = Ti, Zr, Hf, V, Nb, Ta; X = S, Se, Te) form a prototypical family of quasi-1D compounds[10–14] in which strong covalent bonding defines the chain direction, while weaker interchain bonding and van der Waals interactions assemble the chains into sheets and ultimately bulk layered crystals. This unique bonding anisotropy enables TMTCs to combine structural flexibility with highly directional physical behavior, positioning them as promising candidates for next generation nanoelectronics, optoelectronics, and quantum devices. Historically, TMTCs research has centered on classical CDW-bearing systems such as $NbS_3$ and $NbSe_3$, where Peierls instabilities, charge density wave (CDW) order, and pressure- or doping-induced superconductivity underscore the delicate interplay between structure, dimensionality, and electronic ground states[11,15–17]. With the discovery of graphene and the ensuing interest in atomic-scale materials[3,18], quasi-1D vdW structures have gained renewed attention, as reductions in dimensionality often enhance electron–electron interactions and enable unconventional quantum phenomena[1,19]. Indeed, many TMTCs exhibit tunable bandgaps, highly anisotropic electronic dispersion, angle-dependent transport, and strong coupling between structural distortions and electronic instabilities[10,20]. Pressure, an exceptionally clean tuning parameter, has proven particularly powerful in modulating their electronic phases, as demonstrated by pressure-induced suppression of CDW order, reentrant superconductivity in $ZrTe_3$, and structural-electronic reconstructions in layered $TaS_2$ and $TaSe_2$[21–24]. These examples highlight the importance of directly probing how local bonding environments evolve under pressure to drive changes in electronic topology and carrier density.

$CrNbSe_5$ represents an intriguing extension of the TMTC landscape[25,26]. Structurally, it can be viewed as a reconstruction of the transition metal dichalcogenides (TMDC) $CrSe_2$ and TMTC $NbSe_3$, in which the introduction of $3d$ transition metal Cr transforms the $NbSe_3$ chain network into duplex-ladder $[Cr_2Nb_2Se_{10}]$ chains. The combination of potential Cr-based magnetism, quasi-1D chain geometry, and van der Waals layering positions $CrNbSe_5$ as a promising system for studying the coupling between chemical bonding, dimensionality, and physical properties. While its parent compounds display pronounced CDW transitions ($NbSe_3$) or Peierls-like bond instabilities ($CrSe_2$), the structural and electronic behavior of $CrNbSe_5$ remains largely unexplored. Notably, its semimetallic ground state suggests the possibility of rich pressure-tuned quantum phenomena, including metal-insulator transitions, Fermi-surface reconstructions, or emergent superconductivity. A fundamental step toward understanding these possibilities is to clarify how the crystal structure and chemical bonding of $CrNbSe_5$ evolve under external compression. High pressure can modulate orbital overlaps, tune interchain distances, and alter the relative strengths of Cr-Se and Nb-Se interactions, thereby directly controlling the electronic bandwidth and carrier density. However, such effects cannot be reliably inferred from powder diffraction alone, as quasi-1D materials often exhibit subtle distortions, chain sliding, and bond rearrangements that require single-crystal X-ray diffraction (SCXRD) for accurate determination. SCXRD under pressure provides the unique capability to visualize bond-length variations, coordination evolution, and symmetry-preserving structural changes with atomic precision, the information essential for linking structural tuning to emergent physical properties.

The study of bond rearrangements under various pressures has recently been a focus of extensive study in diverse members of the $AT_2X_2$, $AET_2X_2$, and $AAET_4X_4$ families with $A$ = alkali metal, $AE$ = alkali-earth or rare-earth, $T$ = transition metal, and $X$ = pnictogen or chalcogenide. With increasing pressure, the formation of bonds through isostructural phase transitions has led to the creation and/or destruction of magnetic order and moments, remarkable superelastic phenomena, and even superconductivity[27–35]. These prior studies establish bond rearrangement in isostructural transitions as a powerful and general mechanism for tuning physical properties. $CrNbSe_5$ represents a compelling target, as its quasi-one-dimensional bonding framework is expected to be highly susceptible to compression, offering a clean route to induce pronounced changes in its electronic and magnetic ground state.

Here, we combine high-pressure single-crystal X-ray diffraction with complementary physical property measurements to investigate the structural and bonding evolution of $CrNbSe_5$. We reveal how pressure drives continuous modifications of the Cr-Se and Nb-Se polyhedra, enabling reversible transitions between distinct electronic states. By correlating these bonding rearrangements with changes in transport behavior, we demonstrate that $CrNbSe_5$ is a highly responsive quasi-1D system in which electronic properties can be engineered through subtle pressure-induced structural modulation. This work not only establishes the first comprehensive high-pressure structural study of CrNbSe$_5$ but also provides fundamental insight into how chemical bonding controls electronic phase transitions in low-dimensional chalcogenide materials.

**Experimental and Computational Details**

**Crystal Synthesis:** Single crystals of CrNbSe$_5$ were synthesized via a conventional solid-state reaction. Cr pieces (Alfa Aesar, 99% metal basis), Nb slug (Alfa Aesar, 99.95% metal basis, excluding Ta), and Se powder (Thermo Scientific, 99%) were combined in a stoichiometric ratio of 1:1:5 and loaded into an alumina crucible. The crucible was sealed in an evacuated fused-quartz tube under an Ar atmosphere. The sealed tube was heated in a box furnace at a rate of 60 °C/h to 600 °C, held at 600 °C for 3 days, and then furnace-cooled to room temperature. Millimeter-sized single crystals were obtained from the resulting product.

**Ambient Pressure Single Crystal X-ray diffraction (SCXRD):** A CrNbSe$_5$ single crystal (0.14 × 0.02 × 0.01 mm$^3$) was mounted on a nylon loop with Paratone oil and measured at room temperature using a Rigaku XtaLAB Synergy Dualflex diffractometer with a HyPix detector. Data were collected using ω scans with Mo Kα radiation (λ = 0.71073 Å) from a micro-focus sealed tube (50 kV, 1 mA). Data collection strategies were optimized using CrysAlisPro (v1.171.43.120a). Data reduction included Lorentz and polarization corrections, numerical absorption correction based on Gaussian integration over a multifaceted crystal model[36], and empirical spherical harmonics correction using the SCALE3 ABSPACK algorithm[37]. Structures were solved and refined using Olex2 with the SHELXTL package[38–40].

**High-pressure SCXRD:** High-pressure single-crystal XRD was performed at room temperature on the same CrNbSe$_5$ crystal characterized at ambient pressure, using a Diacell One20DAC (Almax-easyLab) with 500 μm culet Boehler-Almax anvils and Rigaku XtaLAB Synergy Dualflex diffractometer with a HyPix detector. A stainless-steel gasket (250 μm thick) was pre-indented to 44 μm, and a 300 μm hole was drilled by the electric discharge machine to hold the sample. A 4:1 methanol-ethanol mixture was used as the pressure-transmitting medium to maintain hydrostatic conditions[41]. Pressures up to 14.6 GPa were determined using ruby R$_1$ fluorescence[42–44]. Refinement details and atomic coordination parameters for ambient pressure and representative pressures (0.8 GPa, 5.9 GPa, and 14.6 GPa) are summarized in **Tables 1** and **2.**

**High-pressure Electrical Resistance Measurements:** High-pressure electrical resistance measurements were conducted using a Be-Cu diamond anvil cell with 300 μm culet anvils in an in-house physical properties measurement system. A T301 stainless-steel gasket (200 μm thick) was pre-indented to 30 μm, and a 280 μm hole was drilled at its center. The gasket surface was insulated with a cubic boron nitride-epoxy mixture. Single-crystal $CrNbSe_5$ samples were measured in a van der Pauw four-probe configuration using Pt foil electrodes. Two independent runs were performed on separate crystals with powdered NaCl as the pressure-transmitting medium. Pressure was applied at room temperature and calibrated by ruby fluorescence[45].

**High-pressure Raman spectroscopy measurements:** High-pressure Raman measurements were performed using a diamond anvil cell with rhenium gaskets, a ruby pressure calibrant, and a 4:1 methanol-ethanol pressure medium. Raman spectra were collected at room temperature up to 40 GPa using a 532 nm laser with a constant power of 40 mW.

**Electronic Structure Calculations:** Density functional theory (DFT) calculations were performed using Quantum ESPRESSO (v7.4.1)[46,47]. Projector-augmented wave (PAW) pseudopotentials from the PSlibrary were employed together with the Perdew-Burke-Ernzerhof (PBE) exchange-correlation functional[48,49]. A kinetic-energy cutoff of 300 Ry was used for the wavefunctions, with the charge-density cutoff set to eight times this value. Brillouin-zone integrations were carried out using a 5 × 11 × 5 Monkhorst-Pack $k$-point mesh [50]. Convergence tests ensured that the total energy varied by less than 1 meV per atom, and the self-consistent-field convergence threshold was set to $10^{-9}$ Ry. High-symmetry $k$-path generation for band-structure calculations was performed using the Spglib library[51,52]. Crystal orbital Hamilton population (COHP) analyses were conducted using LOBSTER (v5.1.1) together with LOPOSTER[53,54]. Fermi surfaces were visualized using FermiSurfer[55].

## Results and Discussion

Low-dimensional solids are often described as stacks of slabs or juxtapositions of one-dimensional building units. CrNbSe$_5$ is such a low-dimensional material, crystallizing in the monoclinic space group $P2_1/m$ at ambient pressure (see **Fig. 1a**). Its structure can be described as $(Cr_2)_{oct}(Nb_2)_{tp}Se_{10}$, consisting of chains of edge-sharing CrSe$_6$ octahedra (oct) and NbSe$_6$ trigonal prisms (tp), characteristic of the FeNb$_3$Se$_{10}$ structure type. This anisotropic connectivity renders the lattice particularly sensitive to external pressure, providing an effective route to tune both structure and physical properties.

To investigate the pressure response, high-pressure SCXRD measurements were performed up to 14.6 GPa. Selected refinement parameters of representative pressures are summarized in **Table 1**. Across the entire pressure range investigated, refinements in the monoclinic $P2_1/m$ space group consistently yielded the best agreement, with no evidence for symmetry lowering. Atomic coordinates refined at ambient pressure and representative pressures of 0.8, 5.9, and 14.6 GPa are listed in **Table 2**. Analysis of the refined structures reveals that the $y$ and $z$ atomic coordinates remain largely invariant with pressure, whereas significant changes in the $x$ coordinates emerge above 5.9 GPa. As illustrated in **Figs 1a-c**, increasing pressure to 5.9 GPa induces a pronounced rearrangement of atomic positions while preserving the overall crystallographic symmetry, marking an iso-symmetric structural transition. This transition is reconstructive in nature, involving a reorganization of local coordination environments rather than a simple, continuous atomic displacement driven by lattice compression. Above 5.9 GPa, the high-pressure phase remains structurally stable up to 14.6 GPa. In this regime, the global framework of the structure is preserved, and further compression primarily manifests as continuous changes in interatomic distances. No additional symmetry breaking or abrupt atomic rearrangements are observed, indicating that the pressure-induced phase is robust once formed. The microscopic origin of the iso-symmetric transition is clarified by examining the evolution of local coordination polyhedra (**Figs 1d-f**). Both the CrSe$_6$ octahedra and NbSe$_6$ trigonal prisms undergo substantial reorientation and changes in polyhedral connectivity during the transition. Following this rearrangement, several Cr-Se and Nb-Se distances shift into typical bonding ranges, consistent with the formation of new pressure-stabilized interactions. The atomic reorganization is further constrained by the emergence of a mirror plane at (1/4, 0, 0), as identified from the refined atomic coordinates in **Table 2**, underscoring the symmetry-preserving nature of the transition.

**Table 1.** Crystal structure and refinement data for CrNbSe$_5$ at 300 K (ambient pressure) and under 0.8 GPa, 5.9 GPa, and 14.6 GPa. Values in parentheses represent estimated standard deviations from the refinements.

| Pressure (GPa) | Ambient | 0.8 | 5.9 | 14.6 |
|---|---|---|---|---|
| Space Group | $P2_1/m$ | $P2_1/m$ | $P2_1/m$ | $P2_1/m$ |
| Unit Cell dimensions | a = 9.1969(9) Å<br>b = 3.5114(2) Å<br>c = 10.4164(9) Å<br>β = 115.52(1)° | a = 9.1080(18) Å<br>b = 3.4938(4) Å<br>c = 10.361(3) Å<br>β = 115.5(3)° | a = 8.7923(13) Å<br>b = 3.4305(3) Å<br>c = 10.0973(13) Å<br>β = 115.44(2)° | a = 8.4490(18) Å<br>b = 3.3508(4) Å<br>c = 9.797(2) Å<br>β = 114.70(2)° |
| Volume | 303.57(5) Å$^3$ | 297.58(13) | 275.02(7) | 252.00(9) |
| Z | 1 | 1 | 1 | 1 |
| Density (calculated) | 5.904 | 6.023 | 6.517 | 7.133 |
| Absorption coefficient | 33.508 | 34.183 | 36.987 | 40.365 |
| F(000) | 470 | 470 | 470.0 | 470 |
| 2θ range | 7.826 to 84.462 | 7.958 to 75.22 | 8.126 to 74.54 | 8.366 to 74.58 |
| Reflections collected | 6972 | 6174 | 3415 | 4260 |
| Independent reflections | 2203 [$R_{int}$ = 0.0734] | 849 [$R_{int}$ = 0.2315] | 854 [$R_{int}$ = 0.0646] | 777 [$R_{int}$ = 0.3230] |
| Data/restraints/parameters | 2203/0/43 | 849/0/44 | 854/0/43 | 777/0/43 |
| Final $R$ indices | $R_1$ (I>2σ(I)) = 0.0509;<br>$wR_2$(I>2σ(I)) = 0.1247<br>$R_1$ (all) = 0.0765;<br>$wR_2$ (all) = 0.1330 | $R_1$ (I>2σ(I)) = 0.0867;<br>$wR_2$(I>2σ(I)) = 0.2190<br>$R_1$ (all) = 0.2013;<br>$wR_2$ (all) = 0.3109 | $R_1$ (I>2σ(I)) = 0.1466;<br>$wR_2$(I>2σ(I)) = 0.3918<br>$R_1$ (all) = 0.2253;<br>$wR_2$ (all) = 0.4970 | $R_1$ (I>2σ(I)) = 0.1809;<br>$wR_2$(I>2σ(I)) = 0.4240<br>$R_1$ (all) = 0.3004;<br>$wR_2$ (all) = 0.5211 |
| Largest diff. peak and hole (e$^-$/Å$^3$) | +4.74 / -2.07 | +3.02 / -4.28 | +7.75 / -13.24 | +6.05 / -7.29 |
| Goodness-of-fit on F$^2$ | 1.046 | 1.010 | 2.073 | 1.571 |

**Table 2.** Atomic coordinates and equivalent isotropic atomic displacement parameters (Å$^2$) for CrNbSe$_5$ at 300 K (ambient pressure) and at 300 K under 0.8 GPa, 5.9 GPa, and 14.6 GPa.

| P (GPa) | Atom | Wyck. | x | y | z | Occ. | $U_{eq}$ |
|---|---|---|---|---|---|---|---|
| Ambient | Nb | 2e | 0.77105(8) | 1/4 | 0.86385(7) | 1 | 0.011(2) |
| | Cr | 2e | 0.06578(14) | 1/4 | 0.40016(14) | 1 | 0.012(2) |
| | Se1 | 2e | 0.46766(9) | 1/4 | 0.25444(8) | 1 | 0.014(15) |
| | Se2 | 2e | 0.33704(9) | 1/4 | 0.00594(8) | 1 | 0.013(15) |
| | Se3 | 2e | 0.72780(9) | 1/4 | 0.58765(8) | 1 | 0.012(14) |
| | Se4 | 2e | 0.15153(10) | 1/4 | 0.65935(9) | 1 | 0.014(15) |
| | Se5 | 2e | 0.01468(9) | 1/4 | 0.13412(7) | 1 | 0.009(13) |
| 0.8 | Nb | 2e | 0.76870(7) | 1/4 | 0.86350(4) | 1 | 0.028(2) |
| | Cr | 2e | 0.06550(12) | 1/4 | 0.40050(7) | 1 | 0.022(2) |
| | Se1 | 2e | 0.47120(8) | 1/4 | 0.25460(5) | 1 | 0.020(2) |
| | Se2 | 2e | 0.33750(9) | 1/4 | 0.00500(5) | 1 | 0.035(2) |
| | Se3 | 2e | 0.72520(8) | 1/4 | 0.58690(5) | 1 | 0.031(2) |

| | Atom | Wyck. | x | y | z | Occ. | U |
|---|---|---|---|---|---|---|---|
| | Se4 | 2e | 0.15290(9) | 1/4 | 0.65990(5) | 1 | 0.029(2) |
| | Se5 | 2e | 0.01680(8) | 1/4 | 0.13510(4) | 1 | 0.022(2) |
| **5.9** | Nb | 2e | 0.60260(8) | 1/4 | 0.86120(6) | 1 | 0.044(3) |
| | Cr | 2e | 0.83970(13) | 1/4 | 0.40240(10) | 1 | 0.029(4) |
| | Se1 | 2e | 0.27460(9) | 1/4 | 0.25870(6) | 1 | 0.029(3) |
| | Se2 | 2e | 0.15620(9) | 1/4 | 0.00200(7) | 1 | 0.037(3) |
| | Se3 | 2e | 0.36690(9) | 1/4 | 0.57970(6) | 1 | 0.039(3) |
| | Se4 | 2e | 0.00770(9) | 1/4 | 0.66730(7) | 1 | 0.041(3) |
| | Se5 | 2e | 0.62130(9) | 1/4 | 0.13660(6) | 1 | 0.043(3) |
| **14.6** | Nb | 2e | 0.60400(10) | 1/4 | 0.86190(8) | 1 | 0.054(3) |
| | Cr | 2e | 0.84030(15) | 1/4 | 0.41260(12) | 1 | 0.039(5) |
| | Se1 | 2e | 0.26890(12) | 1/4 | 0.26610(9) | 1 | 0.060(4) |
| | Se2 | 2e | 0.14470(10) | 1/4 | 0.00140(9) | 1 | 0.048(4) |
| | Se3 | 2e | 0.36690(12) | 1/4 | 0.57540(9) | 1 | 0.066(5) |
| | Se4 | 2e | 0.00440(12) | 1/4 | 0.67490(10) | 1 | 0.059(4) |
| | Se5 | 2e | 0.62860(12) | 1/4 | 0.13850(9) | 1 | 0.049(3) |

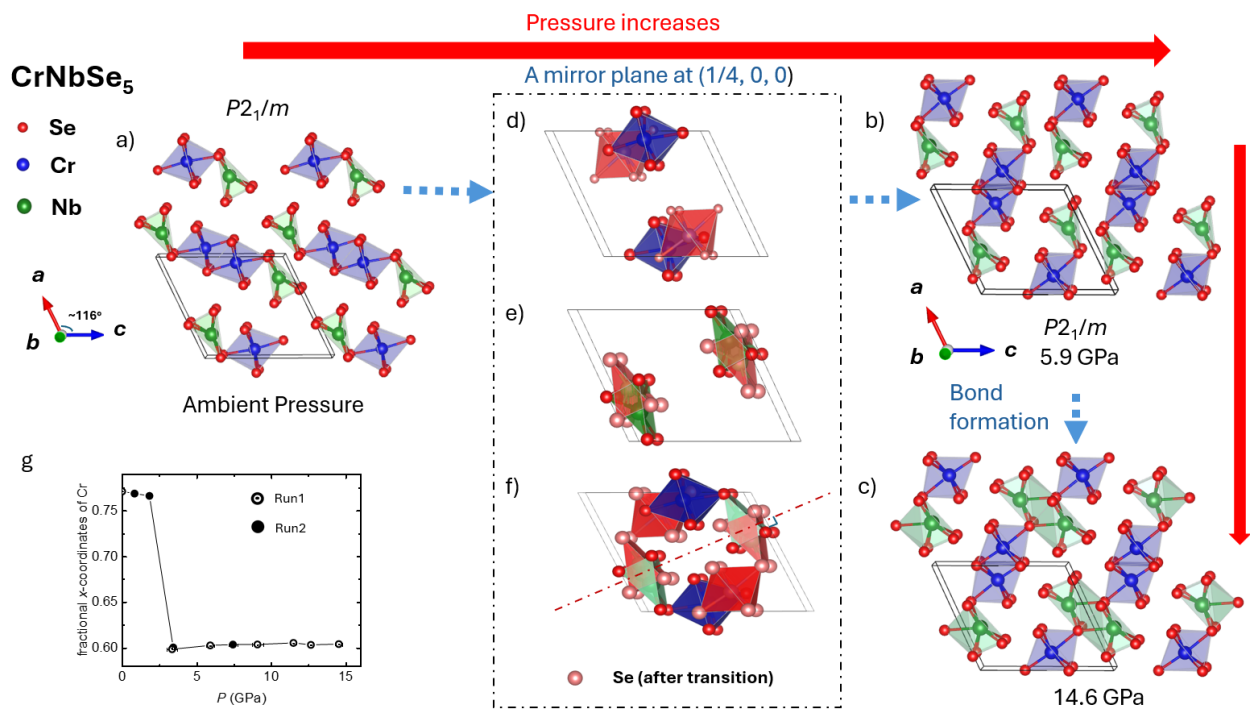

**Fig. 1|** Crystal structure of CrNbSe$_5$ and mirror-plane–mediated iso-symmetric structural transition under pressure. Panels **(a)**-**(c)** show the ambient- and high-pressure $P2_1/m$ crystal structures visualized using VESTA. The red arrow indicates the direction of increasing pressure. Panels **(d)**-**(f)** illustrate the pressure-induced reorganization of atomic positions associated with the iso-symmetric transition; red polyhedra highlight the coordination units undergoing positional rearrangement across the transition. **(g)** shows the fractional x-coordinates of Cr. Solid symbols indicate the data from the sample 2 run 2

While structural refinements reveal a symmetry-preserving yet reconstructive atomic rearrangement, the macroscopic lattice response provides complementary insight into how the framework accommodates compression and signals the emergence of a pressure-induced phase. **Figs 2a-d** summarize the evolution of the lattice parameters as a function of pressure. All lattice parameters decrease monotonically upon compression, with total reductions approaching ~10% at 14.6 GPa, consistent with progressive densification of the low-dimensional framework. The pressure dependence of the unit-cell volume is shown in **Fig 2e** and is well described by a Birch-Murnaghan equation of state, indicating the absence of a first-order volume collapse across the measured pressure range. Despite this smooth volumetric compression, repeated measurements consistently reveal a subtle anomaly near ~3 GPa, most prominently reflected in the $\beta$ angle and $c$-axis parameter in **Fig 2b** and **2d**, respectively. This pressure range, highlighted in violet in **Fig 2e**, marks the formation of a distinct high-pressure structural configuration. Importantly, the structures on both sides of this transition retain a mirror plane at (1/4, 0, 0), an operation that is not an intrinsic symmetry element of the $P2_1/m$ space group but instead emerges as a constraint imposed by the pressure-driven atomic rearrangement. Using a rigid translation (chosen to exactly align the Nb site) and allowing an optimal one to one reassignment of five Se sites, we compared the 0.8 GPa and 3.4 GPa structure. While Nb overlaps exactly and three Se sits can be matched within 0.3 Å, the Cr site still shows a moderate mismatch about 0.6 Å and other two Se sites exhibit large residuals of 1.7 (Se1) and 2 Å (Se4). This is the reason why these structures look similar after rigid rotation, but they cannot be use as the same structure solve mode in SCXRD solve process. This iso-symmetric structural transformation is further corroborated by high-pressure Raman spectroscopy, which exhibits corresponding anomalies in the same pressure range (**Fig S1**), providing independent evidence for pressure-induced coordination reconstruction rather than a conventional symmetry-breaking phase transition.

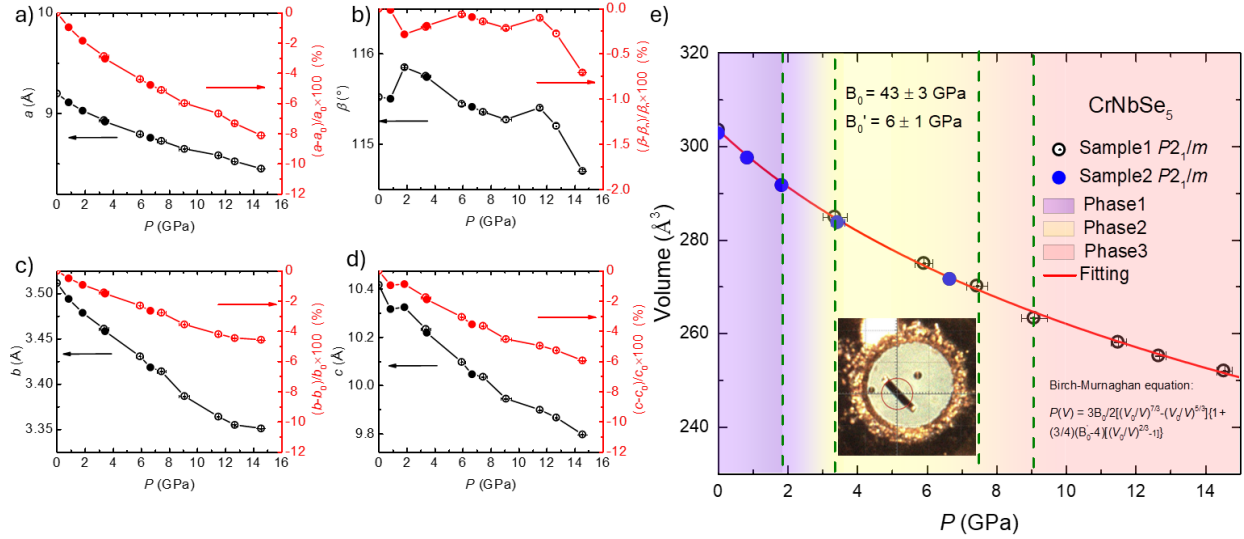

**Fig. 2| Pressure-dependent evolution of the lattice parameters and unit-cell volume of CrNbSe$_5$.** Panels (***a***)-(***d***) show the lattice parameters as a function of pressure; the red panels indicate the relative changes expressed as percentages. Solid symbols indicate the data from sample 2 run 2. (***e***) Unit-cell volume versus pressure with a Birch-Murnaghan equation-of-state fit. The inset displays the gasket and sample chamber at 5.9 GPa; the red circle marks a diameter of 0.1 mm. The light-violet region denotes phase I, corresponding to the ambient-pressure atomic configuration, while the yellow region indicates the pressure-induced phase II with reconstructed atomic positions. The red region indicates the phase III with Fermi surface reconstruction.

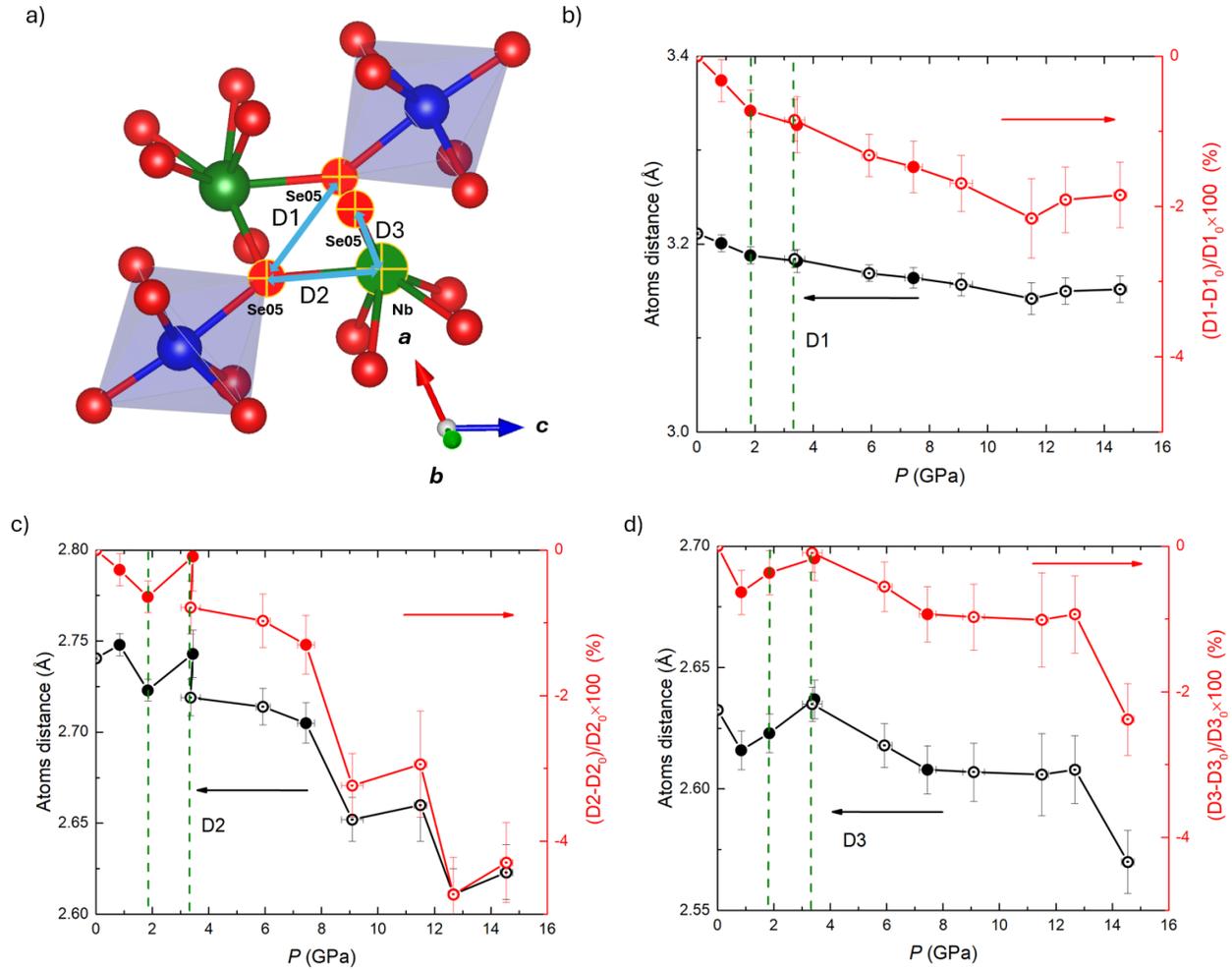

**Fig. 3| Pressure-dependent evolution of selected interatomic distances in CrNbSe$_5$.** (*a*) Schematic illustration of the local coordination environment highlighting the Se5 site (D1), which coordinates to two symmetry-related Nb atoms, giving rise to two distinct Nb-Se distances (D2 and D3). Panels (***b***)-(***d***) show the pressure dependence of these interatomic distances. The red panels indicate the relative changes in bond lengths, expressed as percentages. Solid symbols indicate the data from sample 2 run 2

Although no clear phase transition is evident from the smooth evolution of the unit-cell volume, the pressure dependence of specific Cr-Se and Nb-Se interatomic distances provides deeper insight into subtle structural rearrangements and their influence on electronic properties. **Fig 3** summarizes the evolution of selected bond lengths as a function of pressure, with the green dashed region highlighting the pressure range associated with the iso-symmetric structural transition. Overall, the Cr-Se and Nb-Se bond lengths vary only modestly across the transition, indicating that the pressure-induced transformation primarily involves a reorganization of coordination polyhedra rather than simple atomic displacements.

Repeated measurements show good reproducibility of most bond distances over the entire pressure range, with the notable exception of the Nb-Se bond (D2). The larger uncertainty and enhanced variability of D2 suggest that this interaction plays a key role in accommodating the polyhedral rearrangement near ~3 GPa. In addition, a pronounced change in the D2 bond length is observed around ~8 GPa, closely coinciding with anomalies detected in high-pressure electrical resistance measurements (**Fig S2a**), highlighting a strong coupling between local bonding reconstruction and electronic transport behavior. We suggest that D2 may form the bond like the collapsed tetragonal in 112 system after 8 GPa. Despite these pressure-induced structural and electronic anomalies, no superconductivity was detected up to 55 GPa, which may reflect differences in sample response or pressure conditions. As shown in **Fig S2**, resistance measurements reveal a pressure-driven evolution of the electronic state in $CrNbSe_5$: the material transitions from a semiconducting state to a semimetallic state between ~2.3 GPa and 5 GPa, accompanied by a marked decrease in resistance. Upon further compression, $CrNbSe_5$ reenters a semiconducting behavior at 10.7 GPa (The resistance increases with decreasing temperature.) after exhibiting semimetallic behavior at 7.5 GPa. Notably, the pressures associated with these semiconductor-semimetal-semiconductor transitions correlate closely with changes in the Nb-Se bonding interaction (D2), underscoring the central role of local chemical bonding in governing the reversible electronic phase evolution under pressure.

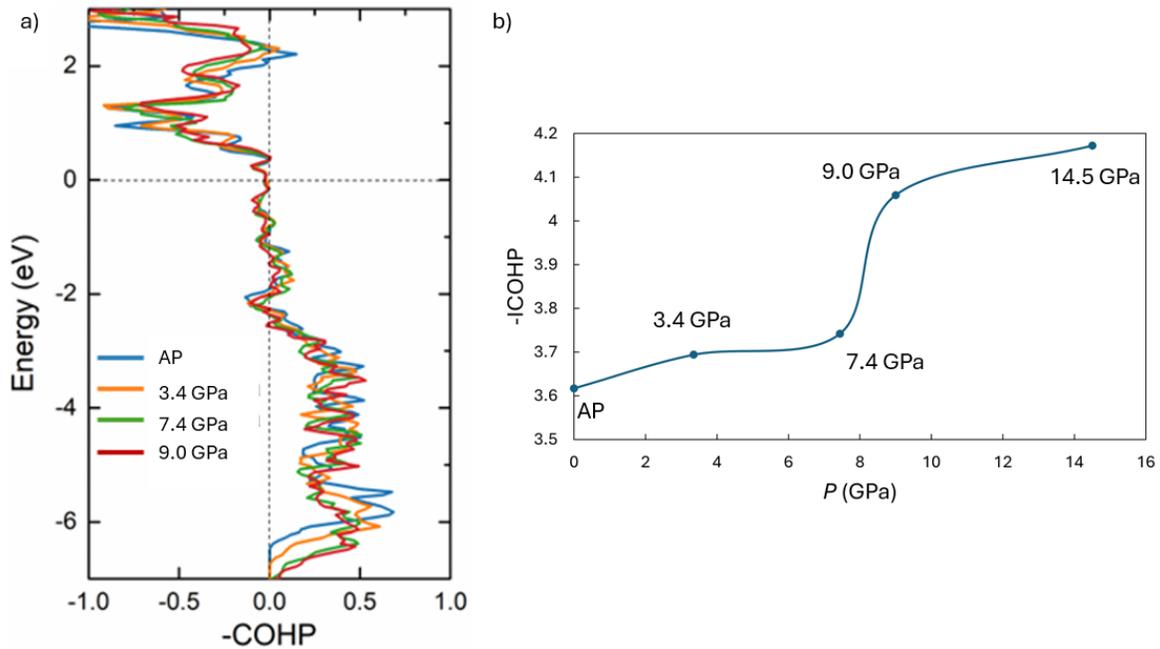

**Fig. 4| Pressure-dependent electronic structure of CrNbSe$_5$.** (*a*) Crystal orbital Hamilton population (COHP) curves for the Nb-Se (D2) bond at selected pressures up to 9 GPa. (***b***) Corresponding integrated COHP (-ICOHP) values for the Nb-Se (D2) interaction as a function of pressure, highlighting the pronounced change above 7.4 GPa.

To elucidate the evolution of chemical bonding and its energetic consequences under pressure, electronic structure calculations were performed for CrNbSe$_5$ at selected pressures. The calculated electronic density of states (DOS) is shown in **Fig S3**. At ambient pressure, CrNbSe$_5$ exhibits a finite DOS at the Fermi level, indicating a metallic electronic structure. Upon compression, the overall DOS broadens, with spectral weight extending to lower energies below −6 eV and higher energies above +2 eV relative to the ambient-pressure case. This bandwidth expansion reflects enhanced orbital overlap induced by pressure. To gain insight into the pressure-dependent chemical bonding, crystal orbital Hamilton population (COHP) analyses and corresponding integrated COHP (-ICOHP) values were evaluated. Consistent with the structural motifs, the dominant orbital interactions arise from Cr-Se and Nb-Se bonds, as shown in **Fig. 4**. Among these, the Nb-Se interaction associated with the D2 bond exhibits the most pronounced pressure sensitivity and is therefore expected to play a critical role in governing the electronic transport behavior. As shown in **Fig. 4a**, the Nb-Se (D2) COHP curve places the Fermi level within a nonbonding to weakly antibonding region, indicating that small pressure-induced changes in bonding can strongly influence carrier behavior. Analysis of the -ICOHP values reveals a significant modification of the Nb-Se (D2) interaction between 7.4 GPa and 9 GPa, in excellent agreement with both the observed structural rearrangement and anomalies in electrical resistance. This behavior can be rationalized within a bonding framework ($S_{ij}^2/\Delta E^{(0)}$) in which pressure enhances orbital overlap (increasing the overlap integral $S_{ij}$), while the energy separation between interacting states ($\Delta E^{(0)}$) remains largely unchanged, leading to strengthened covalent interactions. To further assess the overall bonding evolution, COHP analyses for all relevant atomic pairs (Cr-Cr, Cr-Se, Nb-Nb, Nb-Se, and Se-Se) were performed up to 9 GPa (**Fig. S4**). Notably, a distinct Cr-Nb interaction emerges above ~7.4 GPa and becomes observed at 9 GPa, signaling the onset of new interchain or interpolyhedral coupling under compression. This emergent interaction provides additional support for a pressure-driven reorganization of the bonding network, rather than a simple lattice contraction, and highlights the key role of chemical bonding reconstruction in stabilizing the high-pressure electronic states of CrNbSe$_5$.

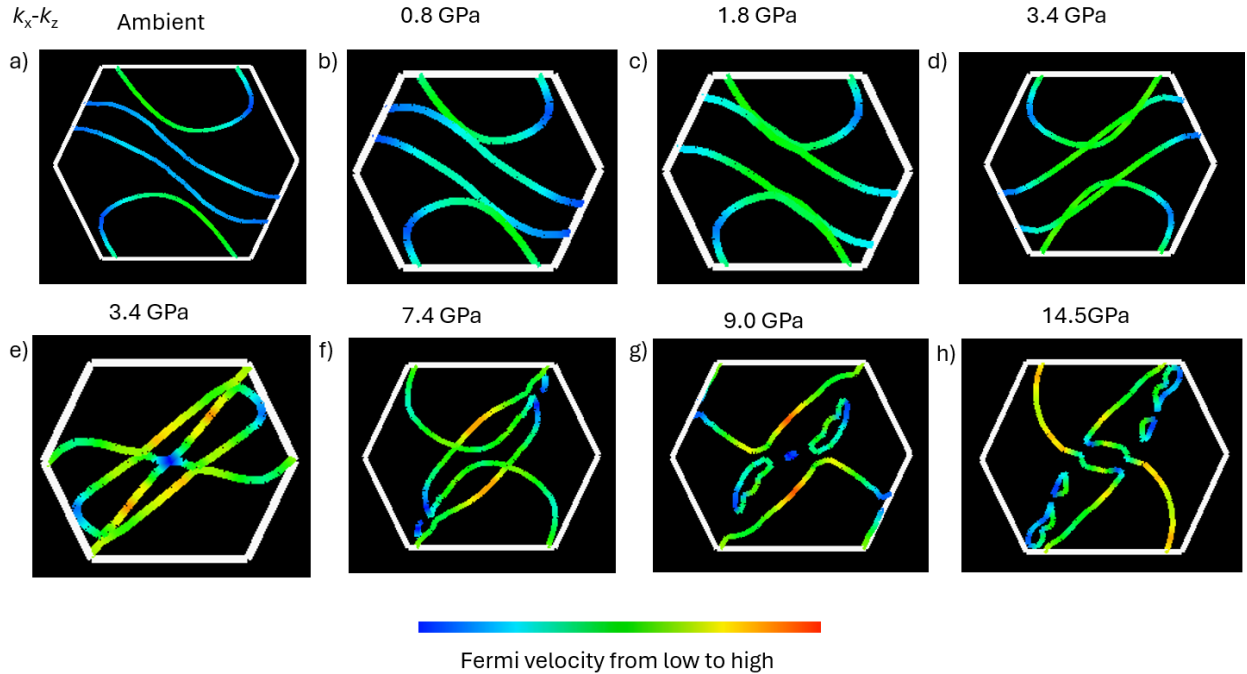

**Fig. 5|** Pressure evolution of the Fermi surface cross sections of CrNbSe$_5$ in the $k_x$-$k_y$ plane and $k_y$-$k_z$ (**Figure S4**). Calculated Fermi surface at (*a*) ambient pressure, (*b*) 0.8 GPa, (*c*) 1.8 GPa, (*d*) and (*e*) 3.4 GPa, (*f*) 7.4 GPa, (*g*) 9.0 GPa, and (*h*) 14.5 GPa is shown within the first Brillouin zone (white hexagon). (*d*) and (*e*) come from the structures of two different measurements. One is around 3.35 GPa (*d*), and another is 3.43 GPa. Considering the precision of pressure measurement using ruby, we marked them as 3.4 GPa for both.

**Fig. 5** gives the evolution of the Fermi surface cross-section of CrNbSe$_5$ in the $k_x$-$k_y$ plane, $k_y$-$k_z$ (**Fig. S5**), and 3D-Fermi surfaces (**Fig. S6**) with the pressure range from ambient to 14.5 GPa. With increasing pressure, the topology of the Fermi surface evolves continuously at low pressures, while a pronounced reconstruction occurs around 8 GPa. In this pressure range, several Fermi surface sheets undergo neck formation and breaking, accompanied by the appearance/disappearance of small electron- and hole-like pockets. These features indicate a pressure-induced Lifshitz transition, i.e., a change in Fermi-surface topology without crystallographic symmetry breaking, consistent with the anomalies observed in transport and bonding analyses. Remarkably, a relatively small pressure between **Fig. 5d** and *e* already produces a change in the D2 bonding distance (**Fig. 3c**), which in turn leads to a pronounced reconstruction of the Fermi surface. Given that other bonding distances exhibit only weak pressure dependence in this regime, the D2 interaction is identified as the key structural parameter governing the pressure-induced electronic reconstruction and the associated changes in the physical properties of $CrNbSe_5$. At higher pressures, however, such as 14.5 GPa, the Fermi surface evolves into a more conventional metallic topology, and no superconducting signature is observed.

**Conclusion**

In summary, applied pressure induces a fully reversible iso-symmetric structural transition in the quasi-one-dimensional compound $CrNbSe_5$ around 3 GPa, accompanied by clear changes in electronic transport without a change in crystallographic symmetry. High-pressure X-ray diffraction, combined with Raman spectroscopy, transport measurements, and electronic structure calculations show that this transition arises from a cooperative reorganization of local Cr-Se and Nb-Se bonding environments and packing method rather than conventional symmetry breaking or dramatic volume collapse. Pressure continuously modifies coordination polyhedra and interchain connectivity, allowing reversible tuning between semiconducting and semimetallic regimes through direct modulation of chemical bonding around 8 GPa. In this pressure range, several Fermi surface sheets undergo neck formation and breaking, accompanied by the emergence and disappearance of small pockets, indicating a change in Fermi surface connectivity. Such a topological transformation of the Fermi surface occurring without any crystallographic symmetry change is characteristic of a Lifshitz transition. This pressure range coincides with pronounced anomalies in the Nb-Se bonding interaction (**Fig. 4**) and the nonmonotonic evolution of electrical resistance (**Fig. S2**), establishing a direct link between pressure-tuned chemical bonding, electronic topology, and electrical transport behavior in $CrNibSe_5$. These results establish $CrNbSe_5$ as a model system in which electronic phase switching is achieved through symmetry-preserving structural reconfiguration and suggest that pressure-controlled bond rearrangement provides a general route for tuning electronic states in low-dimensional chalcogenides.

## Acknowledgments

M.X. thanks Dr. Richard J. Staples for the thoughtful discussion. M.X. and W.X. at Michigan State University, which was supported by the U.S. Department of Energy (DOE), Division of Basic Energy Sciences under Contract DE-SC0023648. C.P. at Weiwei Xie's lab was supported by NSF-DMR-2422361. W.B. acknowledges the support from NSF CAREER Award No. DMR-2045760. The work at Ames National Laboratory was supported by the U.S. Department of Energy, Office of Science, Basic Energy Sciences, Materials Sciences, and Engineering Division. Ames National Laboratory is operated for the U.S. Department of Energy by Iowa State University under contract No. DE-AC02-07CH11358. S.-Y. X. was supported by the NSF Career DMR-2143177.

## References

(1) Balandin, A. A.; Kargar, F.; Salguero, T. T.; Lake, R. K. One-Dimensional van Der Waals Quantum Materials. *Materials Today* **2022**, *55*, 74–91. https://doi.org/10.1016/j.mattod.2022.03.015.

(2) Novoselov, K. S.; Mishchenko, A.; Carvalho, A.; Castro Neto, A. H. 2D Materials and van Der Waals Heterostructures. *Science (1979).* **2016**, *353* (6298). https://doi.org/10.1126/science.aac9439.

(3) Geim, A. K.; Grigorieva, I. V. Van Der Waals Heterostructures. *Nature* **2013**, *499* (7459), 419–425. https://doi.org/10.1038/nature12385.

(4) Katsumata, K. Low-Dimensional Magnetic Materials. *Curr. Opin. Solid State Mater. Sci.* **1997**, *2* (2), 226–230. https://doi.org/10.1016/S1359-0286(97)80070-8.

(5) Mak, K. F.; Shan, J. Photonics and Optoelectronics of 2D Semiconductor Transition Metal Dichalcogenides. *Nat. Photonics* **2016**, *10* (4), 216–226. https://doi.org/10.1038/nphoton.2015.282.

(6) Manzeli, S.; Ovchinnikov, D.; Pasquier, D.; Yazyev, O. V.; Kis, A. 2D Transition Metal Dichalcogenides. *Nat. Rev. Mater.* **2017**, *2* (8), 17033. https://doi.org/10.1038/natrevmats.2017.33.

(7) Yang, S.; Zhang, T.; Jiang, C. Van Der Waals Magnets: Material Family, Detection and Modulation of Magnetism, and Perspective in Spintronics. *Advanced Science* **2021**, *8* (2). https://doi.org/10.1002/advs.202002488.

(8) Wang, Q. H.; Bedoya-Pinto, A.; Blei, M.; Dismukes, A. H.; Hamo, A.; Jenkins, S.; Koperski, M.; Liu, Y.; Sun, Q.-C.; Telford, E. J.; Kim, H. H.; Augustin, M.; Vool, U.; Yin, J.-X.; Li, L. H.; Falin, A.; Dean, C. R.; Casanova, F.; Evans, R. F. L.; Chshiev, M.; Mishchenko, A.; Petrovic, C.; He, R.; Zhao, L.; Tsen, A. W.; Gerardot, B. D.; Brotons-

Gisbert, M.; Guguchia, Z.; Roy, X.; Tongay, S.; Wang, Z.; Hasan, M. Z.; Wrachtrup, J.; Yacoby, A.; Fert, A.; Parkin, S.; Novoselov, K. S.; Dai, P.; Balicas, L.; Santos, E. J. G. The Magnetic Genome of Two-Dimensional van Der Waals Materials. *ACS Nano* **2022**, *16* (5), 6960–7079. https://doi.org/10.1021/acsnano.1c09150.

(9) Li, Y.; Yang, B.; Xu, S.; Huang, B.; Duan, W. Emergent Phenomena in Magnetic Two-Dimensional Materials and van Der Waals Heterostructures. *ACS Appl. Electron. Mater.* **2022**, *4* (7), 3278–3302. https://doi.org/10.1021/acsaelm.2c00419.

(10) Chen, M.; Li, L.; Xu, M.; Li, W.; Zheng, L.; Wang, X. Quasi-One-Dimensional van Der Waals Transition Metal Trichalcogenides. *Research* **2023**, *6*. https://doi.org/10.34133/research.0066.

(11) Ong, N. P.; Monceau, P. Anomalous Transport Properties of a Linear-Chain Metal: $NbSe_3$. *Phys. Rev. B* **1977**, *16* (8), 3443–3455. https://doi.org/10.1103/PhysRevB.16.3443.

(12) Island, J. O.; Barawi, M.; Biele, R.; Almazán, A.; Clamagirand, J. M.; Ares, J. R.; Sánchez, C.; van der Zant, H. S. J.; Álvarez, J. V.; D'Agosta, R.; Ferrer, I. J.; Castellanos-Gomez, A. $TiS_3$ Transistors with Tailored Morphology and Electrical Properties. *Advanced Materials* **2015**, *27* (16), 2595–2601. https://doi.org/10.1002/adma.201405632.

(13) Saleheen, A. I. U.; Chapai, R.; Xing, L.; Nepal, R.; Gong, D.; Gui, X.; Xie, W.; Young, D. P.; Plummer, E. W.; Jin, R. Evidence for Topological Semimetallicity in a Chain-Compound $TaSe_3$. *NPJ Quantum Mater.* **2020**, *5* (1), 53. https://doi.org/10.1038/s41535-020-00257-7.

(14) Chen, F.; Liu, G.; Xiao, Z.; Zhou, H.; Fei, L.; Wan, S.; Liao, X.; Yuan, J.; Zhou, Y. Quasi-One-Dimensional $ZrS_3$ Nanoflakes for Broadband and Polarized Photodetection with High Tuning Flexibility. *ACS Appl. Mater. Interfaces* **2023**, *15* (13), 16999–17008. https://doi.org/10.1021/acsami.3c00273.

(15) Thorne, R. E. Charge-Density-Wave Conductors. *Phys. Today* **1996**, *49* (5), 42–47. https://doi.org/10.1063/1.881498.

(16) Monceau, P.; Peyrard, J.; Richard, J.; Molinié, P. Superconductivity of the Linear Trichalcogenide $NbSe_3$ under Pressure. *Phys. Rev. Lett.* **1977**, *39* (3), 161–164. https://doi.org/10.1103/PhysRevLett.39.161.

(17) Zhu, X.; Ning, W.; Li, L.; Ling, L.; Zhang, R.; Zhang, J.; Wang, K.; Liu, Y.; Pi, L.; Ma, Y.; Du, H.; Tian, M.; Sun, Y.; Petrovic, C.; Zhang, Y. Superconductivity and Charge Density Wave in $ZrTe_{3-x}Se_x$. *Sci. Rep.* **2016**, *6* (1), 26974. https://doi.org/10.1038/srep26974.

(18) Novoselov, K. S.; Geim, A. K.; Morozov, S. V.; Jiang, D.; Zhang, Y.; Dubonos, S. V.; Grigorieva, I. V.; Firsov, A. A. Electric Field Effect in Atomically Thin Carbon Films. *Science (1979).* **2004**, *306* (5696), 666–669. https://doi.org/10.1126/science.1102896.

(19) Voit, J. One-Dimensional Fermi Liquids. *Reports on Progress in Physics* **1995**, *58* (9), 977–1116. https://doi.org/10.1088/0034-4885/58/9/002.

(20) Island, J. O.; Molina-Mendoza, A. J.; Barawi, M.; Biele, R.; Flores, E.; Clamagirand, J. M.; Ares, J. R.; Sánchez, C.; van der Zant, H. S. J.; D'Agosta, R.; Ferrer, I. J.; Castellanos-Gomez, A. Electronics and Optoelectronics of Quasi-1D Layered Transition Metal Trichalcogenides. *2d Mater.* **2017**, *4* (2), 022003. https://doi.org/10.1088/2053-1583/aa6ca6.

(21) Gu, K.; Susilo, R. A.; Ke, F.; Deng, W.; Wang, Y.; Zhang, L.; Xiao, H.; Chen, B. Pressure-Induced Enhancement in the Superconductivity of $ZrTe_3$. *Journal of Physics: Condensed Matter* **2018**, *30* (38), 385701. https://doi.org/10.1088/1361-648X/aada53.

(22) Gleason, S. L.; Gim, Y.; Byrum, T.; Kogar, A.; Abbamonte, P.; Fradkin, E.; MacDougall, G. J.; Van Harlingen, D. J.; Zhu, X.; Petrovic, C.; Cooper, S. L. Structural Contributions to the Pressure-Tuned Charge-Density-Wave to Superconductor Transition in $ZrTe_3$ : Raman Scattering Studies. *Phys. Rev. B* **2015**, *91* (15), 155124. https://doi.org/10.1103/PhysRevB.91.155124.

(23) Grasset, R.; Gallais, Y.; Sacuto, A.; Cazayous, M.; Mañas-Valero, S.; Coronado, E.; Méasson, M.-A. Pressure-Induced Collapse of the Charge Density Wave and Higgs Mode Visibility in 2-$TaS_2$. *Phys. Rev. Lett.* **2019**, *122* (12), 127001. https://doi.org/10.1103/PhysRevLett.122.127001.

(24) Freitas, D. C.; Rodière, P.; Osorio, M. R.; Navarro-Moratalla, E.; Nemes, N. M.; Tissen, V. G.; Cario, L.; Coronado, E.; García-Hernández, M.; Vieira, S.; Núñez-Regueiro, M.; Suderow, H. Strong Enhancement of Superconductivity at High Pressures within the Charge-Density-Wave States of 2H-$TaS_2$ and 2H-$TaSe_2$. *Phys. Rev. B* **2016**, *93* (18), 184512. https://doi.org/10.1103/PhysRevB.93.184512.

(25) Salem, A. B.; Meerschaut, A.; Guemas, L.; Rouxel, J. Characterization of New Low Dimensional Conductors with the $FeNb_3Se_{10}$ Structural Type. *Mater. Res. Bull.* **1982**, *17* (8), 1071–1079. https://doi.org/10.1016/0025-5408(82)90134-9.

(26) Li, C.; Wang, Y.; Li, C.; Liu, K.; Feng, J.; Cheng, H.; Chen, E.; Jiang, D.; Zhang, Q.; Wen, T.; Yue, B.; Yang, W.; Wang, Y. Superconductivity in Quasi-One-Dimensional Antiferromagnetic $CrNbSe_5$ Microwires under High Pressure. *Matter* **2025**, *8* (12), 102299. https://doi.org/10.1016/j.matt.2025.102299.

(27) Ran, S.; Bud'ko, S. L.; Pratt, D. K.; Kreyssig, A.; Kim, M. G.; Kramer, M. J.; Ryan, D. H.; Rowan-Weetaluktuk, W. N.; Furukawa, Y.; Roy, B.; Goldman, A. I.; Canfield, P. C. Stabilization of an Ambient-Pressure Collapsed Tetragonal Phase in $CaFe_2As_2$ and Tuning of the Orthorhombic-Antiferromagnetic Transition Temperature by over 70 K via Control of Nanoscale Precipitates. *Phys. Rev. B* **2011**, *83* (14), 144517. https://doi.org/10.1103/PhysRevB.83.144517.

(28) Jayasekara, W. T.; Kaluarachchi, U. S.; Ueland, B. G.; Pandey, A.; Lee, Y. B.; Taufour, V.; Sapkota, A.; Kothapalli, K.; Sangeetha, N. S.; Fabbris, G.; Veiga, L. S. I.; Feng, Y.; dos Santos, A. M.; Bud'ko, S. L.; Harmon, B. N.; Canfield, P. C.; Johnston, D. C.; Kreyssig, A.; Goldman, A. I. Pressure-Induced Collapsed-Tetragonal Phase in $SrCo_2As_2$. *Phys. Rev. B* **2015**, *92* (22), 224103. https://doi.org/10.1103/PhysRevB.92.224103.

(29) Huyan, S.; Schmidt, J.; Valadkhani, A.; Wang, H.; Li, Z.; Sapkota, A.; Petri, J. L.; Slade, T. J.; Ribeiro, R. A.; Bi, W.; Xie, W.; Mazin, I. I.; Valenti, R.; Bud'ko, S. L.; Canfield, P. C. Near-Room-Temperature Ferromagnetic Ordering in the Pressure-Induced Collapsed-Tetragonal Phase in $SrCo_2P_2$. *Phys. Rev. B* **2025**, *112* (4), L041102. https://doi.org/10.1103/vl33-53qm.

(30) Jia, S.; Williams, A. J.; Stephens, P. W.; Cava, R. J. Lattice Collapse and the Magnetic Phase Diagram of $Sr_{1-x}Ca_xCo_2P_2$. *Phys. Rev. B* **2009**, *80* (16), 165107. https://doi.org/10.1103/PhysRevB.80.165107.

(31) Schmidt, J.; Gorgen-Lesseux, G.; Ribeiro, R. A.; Bud'ko, S. L.; Canfield, P. C. Effects of Co Substitution on the Structural and Magnetic Properties of $Sr(Ni_{1-x}Co_x)_2P_2$. *Phys. Rev. B* **2023**, *108* (17), 174415. https://doi.org/10.1103/PhysRevB.108.174415.

(32) Kaluarachchi, U. S.; Taufour, V.; Sapkota, A.; Borisov, V.; Kong, T.; Meier, W. R.; Kothapalli, K.; Ueland, B. G.; Kreyssig, A.; Valentí, R.; McQueeney, R. J.; Goldman, A. I.; Bud'ko, S. L.; Canfield, P. C. Pressure-Induced Half-Collapsed-Tetragonal Phase in $CaKFe_4As_4$. *Phys. Rev. B* **2017**, *96* (14), 140501. https://doi.org/10.1103/PhysRevB.96.140501.

(33) Song, G.; Borisov, V.; Meier, W. R.; Xu, M.; Dusoe, K. J.; Sypek, J. T.; Valentí, R.; Canfield, P. C.; Lee, S.-W. Ultrahigh Elastically Compressible and Strain-Engineerable Intermetallic Compounds under Uniaxial Mechanical Loading. *APL Mater.* **2019**, *7* (6). https://doi.org/10.1063/1.5087279.

(34) Drachuck, G.; Sapkota, A.; Jayasekara, W. T.; Kothapalli, K.; Bud'ko, S. L.; Goldman, A. I.; Kreyssig, A.; Canfield, P. C. Collapsed Tetragonal Phase Transition in $LaRu_2P_2$. *Phys. Rev. B* **2017**, *96* (18), 184509. https://doi.org/10.1103/PhysRevB.96.184509.

(35) Schmidt, J.; Sapkota, A.; Mueller, C. L.; Xiao, S.; Huyan, S.; Slade, T. J.; Ribeiro, R. A.; Lee, S.-W.; Bud'ko, S. L.; Canfield, P. C. Tuning the Structure and Superconductivity of $SrNi_2P_2$ by Rh Substitution. *Phys. Rev. B* **2025**, *111* (5), 054102. https://doi.org/10.1103/PhysRevB.111.054102.

(36) Parkin, S.; Moezzi, B.; Hope, H. XABS 2: An Empirical Absorption Correction Program. *J. Appl. Crystallogr.* **1995**, *28* (1), 53–56. https://doi.org/10.1107/S0021889894009428.

(37) Walker, N.; Stuart, D. An Empirical Method for Correcting Diffractometer Data for Absorption Effects. *Acta Crystallogr. A* **1983**, *39* (1), 158–166. https://doi.org/10.1107/S0108767383000252.

(38) Sheldrick, G. M. SHELXT – Integrated Space-Group and Crystal-Structure Determination. *Acta Crystallogr. A Found. Adv.* **2015**, *71* (1), 3–8. https://doi.org/10.1107/S2053273314026370.

(39) Sheldrick, G. M. Crystal Structure Refinement with SHELXL. *Acta Crystallogr. C Struct. Chem.* **2015**, *71* (1), 3–8. https://doi.org/10.1107/S2053229614024218.

(40) Dolomanov, O. V.; Bourhis, L. J.; Gildea, R. J.; Howard, J. A. K.; Puschmann, H. OLEX2 : A Complete Structure Solution, Refinement and Analysis Program. *J. Appl. Crystallogr.* **2009**, *42* (2), 339–341. https://doi.org/10.1107/S0021889808042726.

(41) Chen, X.; Lou, H.; Zeng, Z.; Cheng, B.; Zhang, X.; Liu, Y.; Xu, D.; Yang, K.; Zeng, Q. Structural Transitions of 4:1 Methanol–Ethanol Mixture and Silicone Oil under High Pressure. *Matter and Radiation at Extremes* **2021**, *6* (3). https://doi.org/10.1063/5.0044893.

(42) Dewaele, A.; Torrent, M.; Loubeyre, P.; Mezouar, M. Compression Curves of Transition Metals in the Mbar Range: Experiments and Projector Augmented-Wave Calculations. *Phys. Rev. B* **2008**, *78* (10), 104102. https://doi.org/10.1103/PhysRevB.78.104102.

(43) Mao, H. K.; Xu, J.; Bell, P. M. Calibration of the Ruby Pressure Gauge to 800 Kbar under Quasi-hydrostatic Conditions. *J. Geophys. Res. Solid Earth* **1986**, *91* (B5), 4673–4676. https://doi.org/10.1029/JB091iB05p04673.

(44) Piermarini, G. J.; Block, S.; Barnett, J. D.; Forman, R. A. Calibration of the Pressure Dependence of the R1 Ruby Fluorescence Line to 195 Kbar. *J. Appl. Phys.* **1975**, *46* (6), 2774–2780. https://doi.org/10.1063/1.321957.

(45) Shen, G.; Wang, Y.; Dewaele, A.; Wu, C.; Fratanduono, D. E.; Eggert, J.; Klotz, S.; Dziubek, K. F.; Loubeyre, P.; Fat'yanov, O. V.; Asimow, P. D.; Mashimo, T.; Wentzcovitch, R. M. M. Toward an International Practical Pressure Scale: A Proposal for an IPPS Ruby Gauge (IPPS-Ruby2020). *High Press. Res.* **2020**, *40* (3), 299–314. https://doi.org/10.1080/08957959.2020.1791107.

(46) Giannozzi, P.; Baroni, S.; Bonini, N.; Calandra, M.; Car, R.; Cavazzoni, C.; Ceresoli, D.; Chiarotti, G. L.; Cococcioni, M.; Dabo, I.; Dal Corso, A.; de Gironcoli, S.; Fabris, S.; Fratesi, G.; Gebauer, R.; Gerstmann, U.; Gougoussis, C.; Kokalj, A.; Lazzeri, M.; Martin-Samos, L.; Marzari, N.; Mauri, F.; Mazzarello, R.; Paolini, S.; Pasquarello, A.; Paulatto, L.; Sbraccia, C.; Scandolo, S.; Sclauzero, G.; Seitsonen, A. P.; Smogunov, A.; Umari, P.; Wentzcovitch, R. M. QUANTUM ESPRESSO: A Modular and Open-Source Software Project for Quantum Simulations of Materials. *Journal of Physics: Condensed Matter* **2009**, *21* (39), 395502. https://doi.org/10.1088/0953-8984/21/39/395502.

(47) Giannozzi, P.; Andreussi, O.; Brumme, T.; Bunau, O.; Buongiorno Nardelli, M.; Calandra, M.; Car, R.; Cavazzoni, C.; Ceresoli, D.; Cococcioni, M.; Colonna, N.; Carnimeo, I.; Dal Corso, A.; de Gironcoli, S.; Delugas, P.; DiStasio, R. A.; Ferretti, A.; Floris, A.; Fratesi, G.; Fugallo, G.; Gebauer, R.; Gerstmann, U.; Giustino, F.; Gorni, T.; Jia, J.; Kawamura,

M.; Ko, H.-Y.; Kokalj, A.; Küçükbenli, E.; Lazzeri, M.; Marsili, M.; Marzari, N.; Mauri, F.; Nguyen, N. L.; Nguyen, H.-V.; Otero-de-la-Roza, A.; Paulatto, L.; Poncé, S.; Rocca, D.; Sabatini, R.; Santra, B.; Schlipf, M.; Seitsonen, A. P.; Smogunov, A.; Timrov, I.; Thonhauser, T.; Umari, P.; Vast, N.; Wu, X.; Baroni, S. Advanced Capabilities for Materials Modelling with Quantum ESPRESSO. *Journal of Physics: Condensed Matter* **2017**, *29* (46), 465901. https://doi.org/10.1088/1361-648X/aa8f79.

(48) Dal Corso, A. Pseudopotentials Periodic Table: From H to Pu. *Comput. Mater. Sci.* **2014**, *95*, 337–350. https://doi.org/10.1016/j.commatsci.2014.07.043.

(49) Perdew, J. P.; Burke, K.; Ernzerhof, M. Generalized Gradient Approximation Made Simple. *Phys. Rev. Lett.* **1996**, *77* (18), 3865–3868. https://doi.org/10.1103/PhysRevLett.77.3865.

(50) Monkhorst, H. J.; Pack, J. D. Special Points for Brillouin-Zone Integrations. *Phys. Rev. B* **1976**, *13* (12), 5188–5192. https://doi.org/10.1103/PhysRevB.13.5188.

(51) Setyawan, W.; Curtarolo, S. High-Throughput Electronic Band Structure Calculations: Challenges and Tools. *Comput. Mater. Sci.* **2010**, *49* (2), 299–312. https://doi.org/10.1016/j.commatsci.2010.05.010.

(52) Togo, A.; Shinohara, K.; Tanaka, I. Spglib: A Software Library for Crystal Symmetry Search. *Science and Technology of Advanced Materials: Methods* **2024**, *4* (1). https://doi.org/10.1080/27660400.2024.2384822.

(53) Nelson, R.; Ertural, C.; George, J.; Deringer, V. L.; Hautier, G.; Dronskowski, R. LOBSTER : Local Orbital Projections, Atomic Charges, and Chemical-bonding Analysis from Projector-augmented-wave-based Density-functional Theory. *J. Comput. Chem.* **2020**, *41* (21), 1931–1940. https://doi.org/10.1002/jcc.26353.

(54) Wang, Y.; Müller, P. C.; Hemker, D.; Dronskowski, R. LOPOSTER: A Cascading Postprocessor for LOBSTER. *J. Comput. Chem.* **2025**, *46* (17). https://doi.org/10.1002/jcc.70167.

(55) Kawamura, M. FermiSurfer: Fermi-Surface Viewer Providing Multiple Representation Schemes. *Comput. Phys. Commun.* **2019**, *239*, 197–203. https://doi.org/10.1016/j.cpc.2019.01.017.

# Supplementary Information

# Observation of Iso-Symmetric Structural and Lifshitz Transitions in Quasi-one-dimensional CrNbSe$_5$

Mingyu Xu[1], Peng Cheng[1], Shuyuan Huyan[2,3], Wenli Bi[4], Su-Yang Xu[5], Sergey L. Bud'ko[2,3], Paul C. Canfield[2,3], Weiwei Xie[1*]

6. Department of Chemistry, Michigan State University, East Lansing, MI 48824, USA
7. Ames National Laboratory, Ames, IA 50011, USA
8. Department of Physics and Astronomy, Iowa State University, Ames, IA 50011, USA
9. SmartState Center for Experimental Nanoscale Physics, Department of Physics and Astronomy, University of South Carolina, Columbia, SC 29208, USA
10. Department of Chemistry and Chemical Biology, Harvard University, Cambridge, MA 02138, USA

[*]Corresponding author: Dr. Weiwei Xie (xieweiwe@msu.edu)

# Table of Contents

**Figure 1.** (*a*) Raman measurements of CrNbSe$_5$ around the pressure reading. (*b*) Raman shift as a function of pressure.

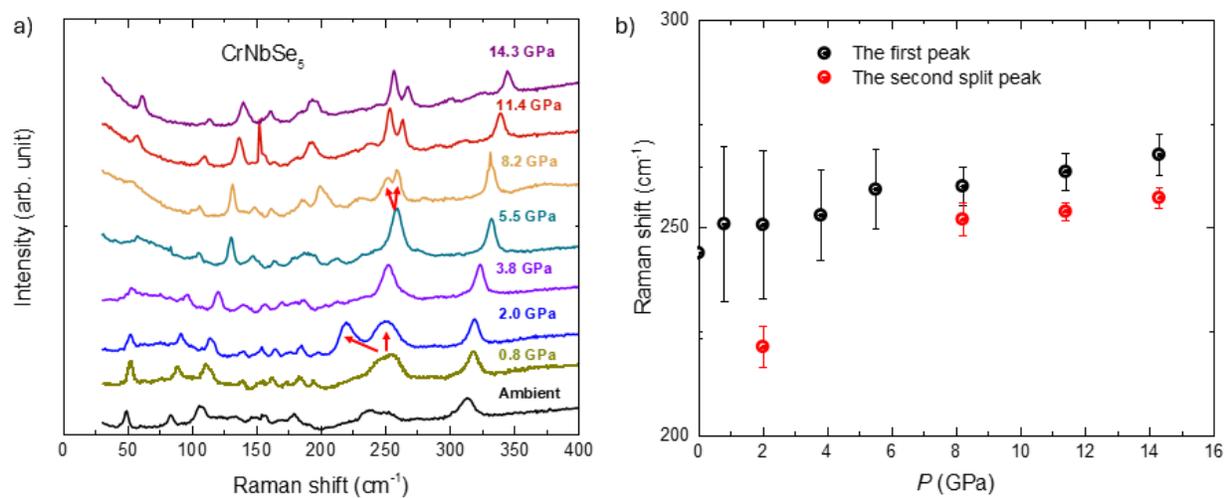

**Figure S2. High-pressure resistance of CrNbSe$_5$. Blue dashed lines indicate the transitions around 3 and 8 GPa.**

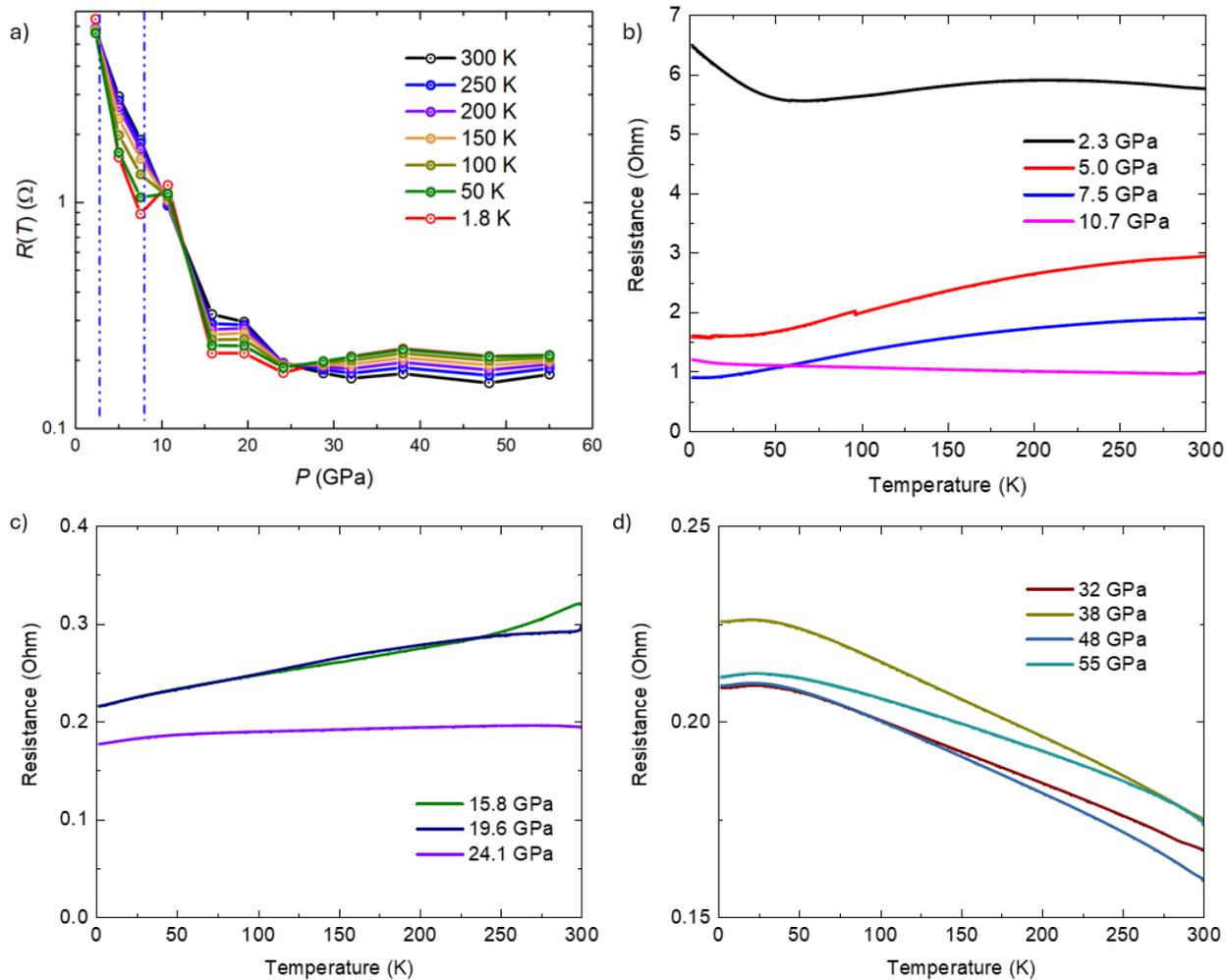

**Figure S3. The calculated electronic density of states (DOS).**

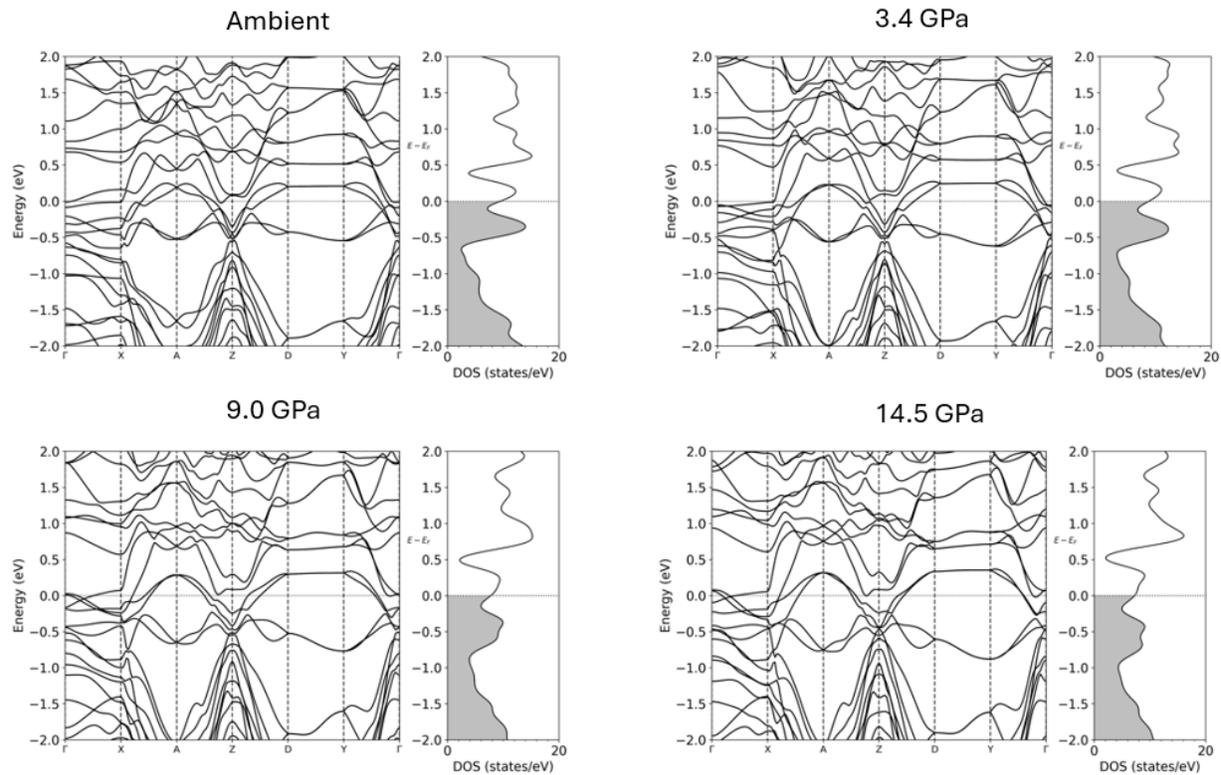

**Figure S4.** Pressure evolution of crystal orbital Hamilton population (COHP) curves for all relevant atomic interactions in CrNbSe$_5$, including Cr-Cr, Cr-Se, Nb-Nb, Nb-Se, and Se-Se. Panels show results at (*a*) ambient pressure, (*b*) 3.4 GPa, (*c*) 7.4 GPa, and (*d*) 9.0 GPa. Notably, a Cr-Nb interaction emerges at high pressure, becoming evident at 9 GPa.

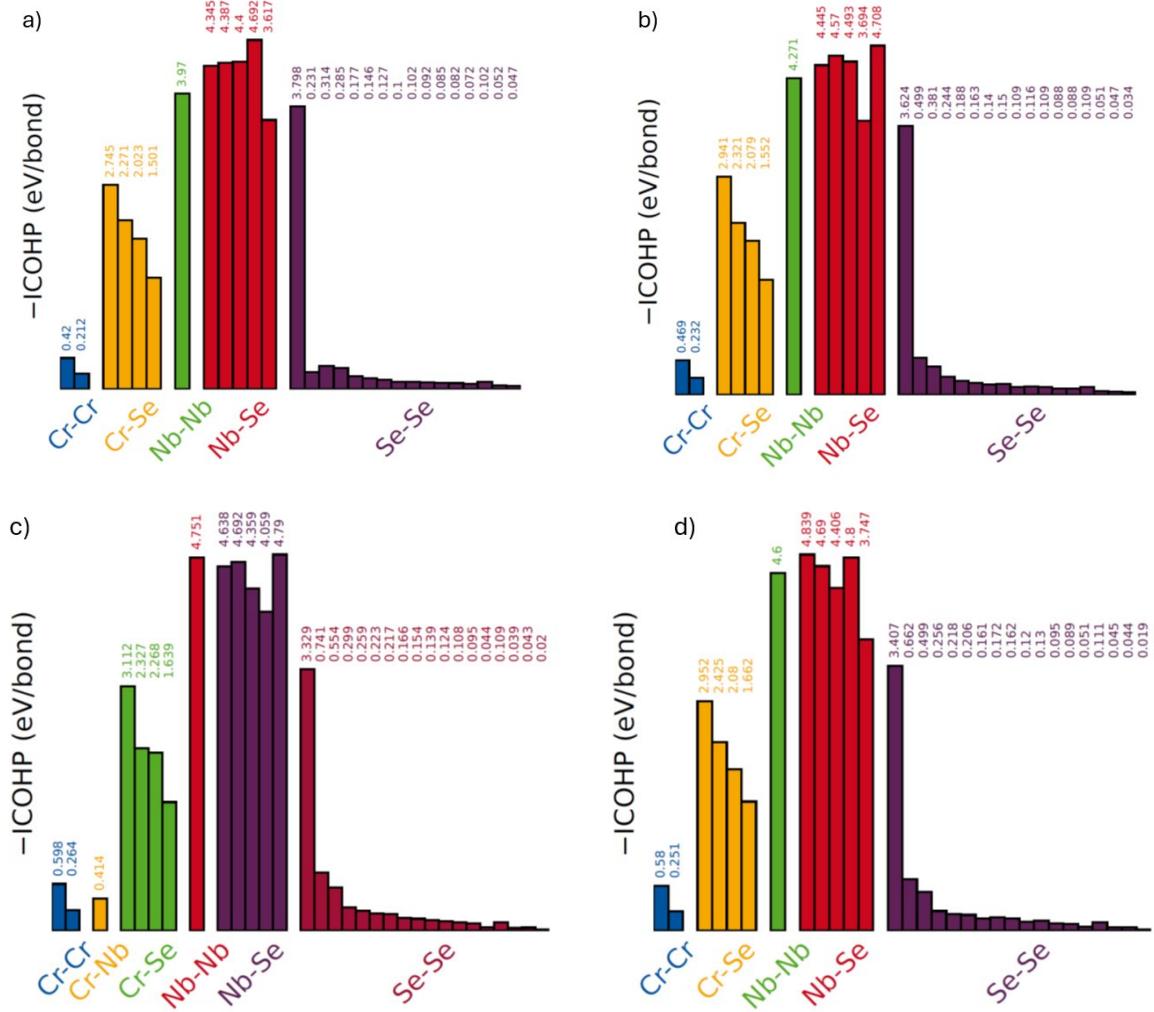

**Figure S5 Pressure evolution of the Fermi surface cross section of CrNbSe$_5$ in the $k_y$-$k_z$ .**

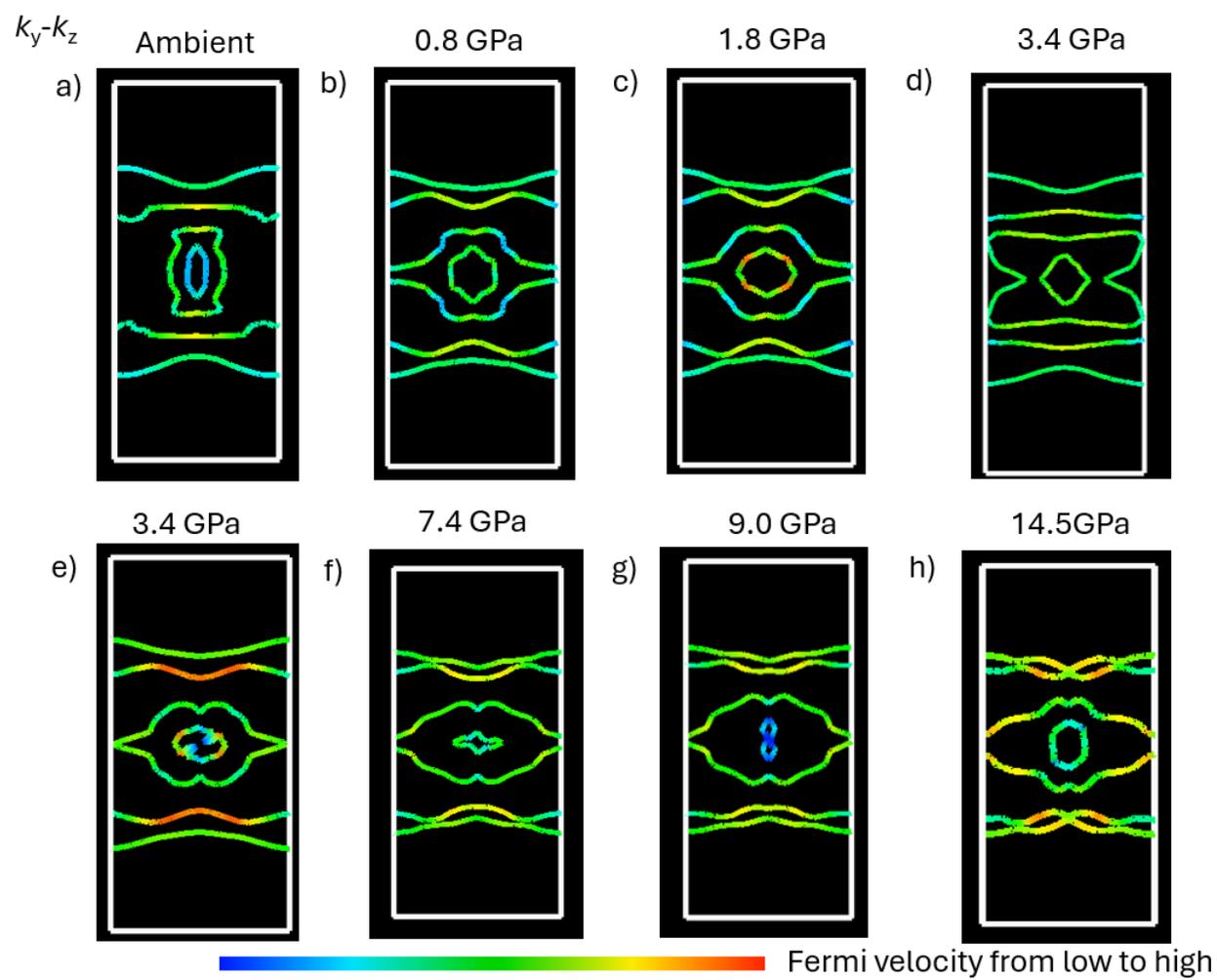

**Figure S6 Pressure evolution of the Fermi surface cross section of CrNbSe$_5$.**

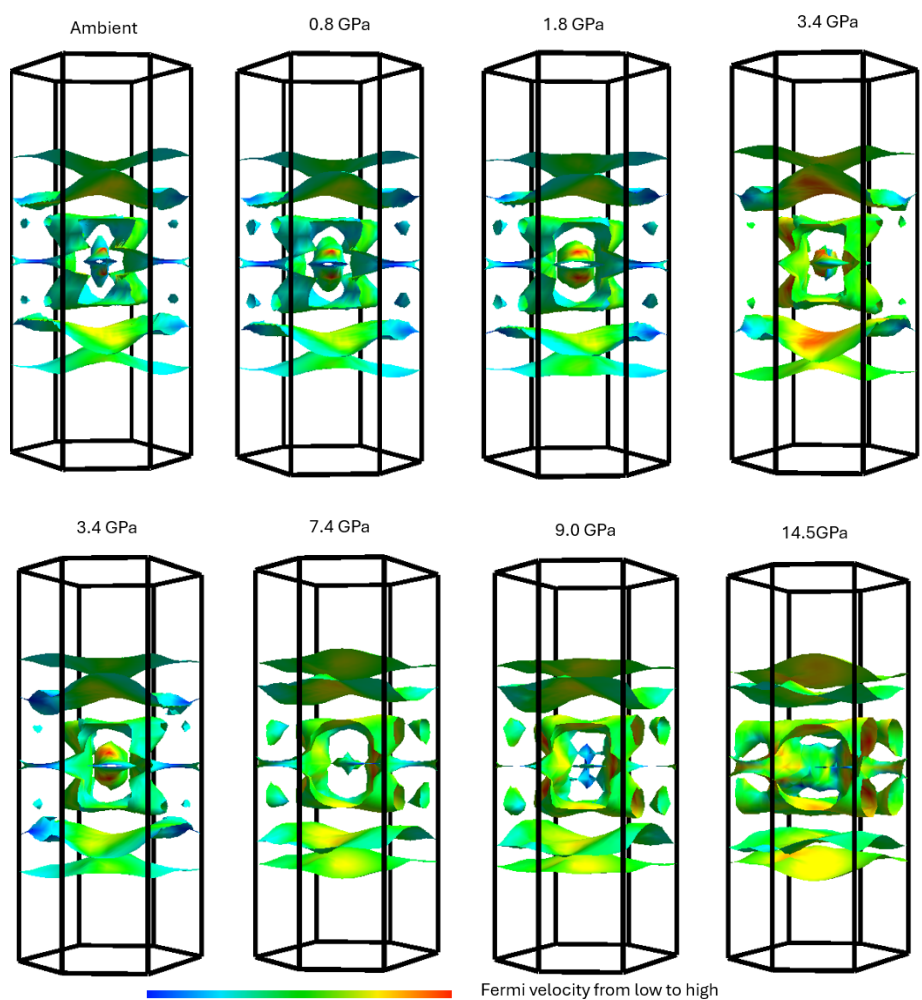